\documentclass[fleqn,10pt]{wlscirep}

\usepackage{multirow}
\usepackage{rotating}
\usepackage{array}
\newcolumntype{L}[1]{>{\raggedright\let\newline\\\arraybackslash\hspace{0pt}}m{#1}}
\newcolumntype{C}[1]{>{\centering\let\newline\\\arraybackslash\hspace{0pt}}m{#1}}
\newcolumntype{R}[1]{>{\raggedleft\let\newline\\\arraybackslash\hspace{0pt}}m{#1}}

\title{Amoeboid-mesenchymal migration plasticity promotes invasion only in complex heterogeneous microenvironments}

\author[1,*]{Katrin Talkenberger}
\author[2,3]{Elisabetta Ada Cavalcanti-Adam}
\author[1]{Andreas Deutsch}
\author[4,1]{Anja Voss-B{\"o}hme}
\affil[1]{Center for Information Services and High Performance Computing, Technische Universit{\"a}t Dresden, 01062 Dresden, Germany}
\affil[2]{Department of Biophysical Chemistry, Institute of Physical Chemistry, Heidelberg University, 69120 Heidelberg, Germany}
\affil[3]{Max Planck Institute for Medical Research, Department of Cellular Biophysics, 69120 Heidelberg, Germany}
\affil[4]{Hochschule f{\"u}r Technik und Wirtschaft Dresden, Friedrich-List-Platz 1, 01069 Dresden}

\affil[*]{katrin.boettger@tu-dresden.de}

\begin{abstract}
During tissue invasion individual tumor cells exhibit two interconvertible migration modes, namely mesenchymal and amoeboid migration. 
The cellular microenvironment triggers the switch between both modes, thereby allowing adaptation to dynamic conditions. 
It is, however, unclear if this amoeboid-mesenchymal migration plasticity contributes to a more effective tumor invasion. 
We address this question with a mathematical model, where the amoeboid-mesenchymal migration plasticity is regulated in response to local extracellular matrix resistance. 
Our numerical analysis reveals that extracellular matrix structure and presence of a chemotactic gradient are key determinants of the model behavior. 
Only in complex microenvironments, if the extracellular matrix is highly heterogeneous and a chemotactic gradient directs migration, the amoeboid-mesenchymal migration plasticity allows a more widespread invasion compared to the non-switching amoeboid and mesenchymal modes. 
Importantly, these specific conditions are characteristic for \emph{in vivo} tumor invasion.
Thus, our study suggests that \emph{in vitro} systems aiming at unraveling the underlying molecular mechanisms of tumor invasion should take into account the complexity of the microenvironment by considering the combined effects of structural heterogeneities and chemical gradients on cell migration.
\end{abstract}
\begin{document}

\flushbottom
\maketitle
%
%
\thispagestyle{empty}

\section*{Introduction}

Solid tumors become invasive if cells migrate away from their initial primary location. 
The tumor cell microenvironment with its variety of biomechanical and molecular cues plays a critical role in the localized invasion throughout the tissue. 
For example, tumor cells are known to react to soluble factors, such as chemokines and growth factors, by directional movement towards the extracellular gradient of chemicals \cite{Roussos2011}. 
The importance of the extracellular matrix (ECM) in tumor invasion has recently received particular attention \cite{Lu2012,Taddei2013}.
The ECM, which fills the space between cells through a complex organization of proteins and polysaccharides, imposes a biomechanical resistance that moving cells need to overcome. To migrate, tumor cells might either degrade the ECM to pass through, or modify their shape and squeeze through the ECM pores \cite{Brabek2010}. These two distinct migration modes are commonly termed "path-generating" mesenchymal and "path-finding" amoeboid mode \cite{Pankova2010,Hecht2015}. The mesenchymal migration mode is characterized by an elongated cell morphology, adherence to the surrounding ECM mediated by integrins and ECM degradation by proteases \cite{Friedl2010}. In contrast, during amoeboid migration, cells are highly deformable, their adhesion to the ECM is rather weak, and proteolytic activity is reduced or absent. The low adhesion of cells in the amoeboid migration mode enables the cells to move comparatively faster than those migrating in mesenchymal migration mode \cite{Liu2015,Pankova2010}. Remarkably, tumor cells are able to adapt their migration mode to changing microenvironmental conditions \cite{Brabek2010,Friedl2010,Friedl2011,Taddei2013}, a feature called migration plasticity. In particular, it has been observed that ECM parameters like density or stiffness, regulate the transition between amoeboid and mesenchymal migration modes, which is very dynamic and comprises intermediate states, where cells display properties of both migratory phenotypes \cite{Friedl2011,Taddei2013,Huttenlocher2011}. At the subcellular to cellular level, the impact of ECM properties on molecular mechanisms of individual cell motility has been studied using both experimental \cite{Friedl2010,Wolf2013} and theoretical \cite{Hecht2011,Sakamoto2011,Tozluoglu2013,Tozluoglu2015} approaches. However, it remains unclear how the adaptation responses of amoeboid and mesenchymal migration modes contribute to the tumor invasion process. In particular, it is not known if and how amoeboid-mesenchymal plasticity allows a more effective invasion compared to the non-adaptive amoeboid or mesenchymal modes. So far, only the impact of interactions between non-switching moving cells and the ECM on tumor invasion has been studied \cite{Brabek2010,Polacheck2013,Hecht2015}. Hecht et al.\ \cite{Hecht2015} considered an agent-based model for fixed proportions of amoeboid and mesenchymal cells migrating in a maze-like environment with directional cue, and showed that an amoeboid cell population can benefit from the presence of a small number of path creating mesenchymal cells. 

In this work, we develop a mathematical model to study the consequences of amoeboid-mesenchymal migration plasticity on tumor invasion in a switching population of cells that adopt their migration mode in response to the ECM conditions. We are interested in the question under what environmental conditions the plasticity of amoeboid and mesenchymal migration modes provides an advantage for tumor invasion. We formulate and analyze a cellular automaton model, where cells are able to switch between a slow mesenchymal migration mode with ECM degradation, and a fast amoeboid migration mode without ECM degradation, depending on the biomechanical resistance imposed by the ECM. 
With computer simulations of the mathematical model, we compare the invasion behavior of a population of cells where each cell is allowed to switch between ameoboid and mesenchymal migration modes with the behavior of a non-switching cell population, under different environmental conditions. In particular, we distinguished spatially homogeneous and heterogeneous ECM resistance conditions, ranging from weakly structured to highly structured, in combination with different responses towards a chemotactic gradient which directs cell migration.
Our numerical analysis reveals that ECM structure and presence of a chemotactic gradient are key determinants of the model behavior. The amoeboid-mesenchymal migration plasticity leads to a more widespread invasion compared to the non-switching situation only in complex microenvironments, more specifically, if the ECM is strongly heterogeneous and a chemotactic gradient directs migration. The latter conditions are characteristic for \emph{in vivo} tumor invasion. This suggests that experimental studies on tumor invasion should represent this complexity of the microenvironment.

\section*{Methods}

\subsection*{The model}

We develop a mathematical model to study the effects of amoeboid-mesenchymal migration plasticity on tumor invasion. To determine the specific impact of migration plasticity of individual cells on overall cell population invasion dynamics, we coarse-grain to a cell-based model, namely a cellular automaton (CA), which is analyzed at the population level.
Cellular automata are a class of spatially and temporally discrete mathematical models which allow to (i) model cell-cell and cell-ECM interactions, as well as cell migration, and (ii) to analyze emergent behavior at the cell population level \cite{Boettger2012,Boettger2015,Deutsch2005,Hatzikirou2008,Hatzikirou2015,Peruani2011,VossBoehme2010}.

We consider the ECM as a physical barrier which imposes a resistance against the moving cell body. A widely studied parameter which mechanically impedes cell movement is the ECM network density. Other physical properties of the ECM, such as porosity, as well as biomechanical properties like ECM stiffness, have been observed to either enable or restrict cell migration. Importantly, the different ECM properties are not independent but rather connected \cite{Lu2012}. Thus, for instance, the density of fibrillar ECM is positively interconnected with stiffness and inversely proportional to pore size, such that alterations of either property impact the overall ECM structure \cite{Wolf2013}. In view of this, we incorporate ECM properties like density and stiffness into a lumped parameter called ``resistance''.

We distinguish two migration modes, namely amoeboid-like (\textit{A}) and mesenchymal-like (\textit{M}) migration. We assume that amoeboid-like cells migrate in a  protease- and integrin-independent way, driven by Rho/ROCK-mediated contractility. In contrast, mesenchymal-like cells migrate protease- and integrin-dependently, where Rho/ROCK signaling is inhibited. Thus, the amoeboid and mesenchymal migration phenotypes in our model represent two ends in a broad spectrum of individual cell migration modes \cite{Huttenlocher2011}. 

We assume that cells change their migration phenotype in dependence of local ECM resistance. This assumption is supported by several studies. In particular, it has been reported that under soft ECM conditions, cells are round-shaped and their migration is independent of protease activity but rather driven by Rho/ROCK-mediated contractility. In contrast, under higher ECM stiffness conditions, Rho/ROCK signaling is inhibited, cells show elongated morphology and their migration relies on protease activity \cite{Brabek2010}. In another recent study it has been shown that in the absence of focal adhesions and under conditions of confinement, mesenchymal cells can spontaneously switch to a fast amoeboid migration phenotype \cite{Liu2015}.
Accordingly, we assume in our model that amoeboid-like cells move relatively fast and are not able to degrade ECM, while mesenchymal-like cells move slower and locally degrade the ECM. We further assume that cell migration speed is highest for low ECM resistance but decreases with increasing ECM resistance. This assumption is based on findings by Wolf et al. \cite{Wolf2013}: migration speed gradually diminishes as a function of decreasing pore size (which is associated with increasing ECM resistance) with a steeper slope when cells migrate in an amoeboid fashion. An optimal migration speed has been observed during migration through tissue of sufficient to high porosity (which corresponds to low ECM resistance). In our coarse-grained model, we do not explicitly account for cell shape changes or Rho/ROCK- and integrin-dependent signaling pathways.

Furthermore, we assume that the direction of cell movement is determined by a constant chemotactic gradient. Cells respond to this external gradient with a certain intensity, independent of their migration phenotype.

Our CA model is described  on a two-dimensional regular lattice $S\subset\mathbb{Z}^2$.  Each lattice site is in one of a set of states $(\eta,\mu)\in\{0,1,2\}\times[0,1]$, with $\eta$ denoting the cell state value and $\mu$ the ECM state value. We interpret the cell state value $0$ as an unoccupied lattice site, cell state value $1$ as an \textit{A}-cell and cell state value $2$ as an \textit{M}-cell. The ECM state value represents the physical resistance of the ECM. 

\begin{figure}[ht]
\centering
\includegraphics[width=\linewidth]{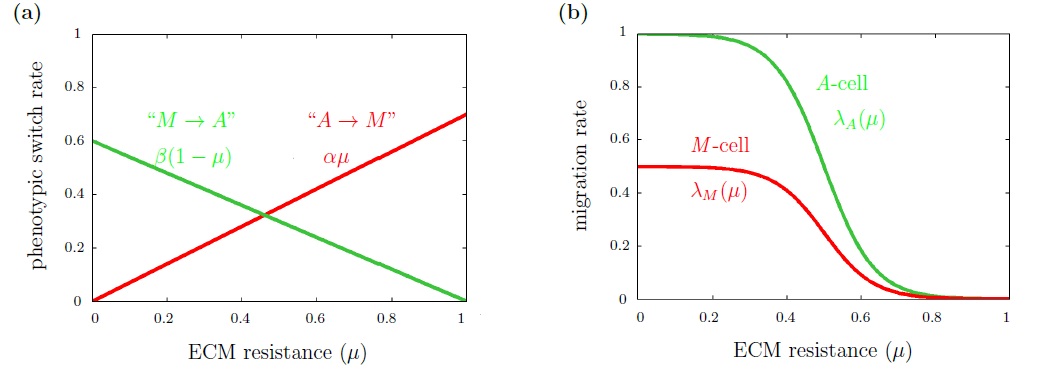}
\caption{Phenotypic switch rates and migration rates as functions of ECM resistance. 
	(a) Phenotypic switch rate. \textit{A}-cells change their phenotype with rate $\alpha\mu$ and \textit{M}-cells with rate $\beta(1-\mu)$. In the plot, the switch parameters are $\alpha=0.7$ and $\beta=0.6$. 
	(b) Migration rate. For \textit{A}-cells, the migration rate is given by equation \eqref{eq:migrationFun-A}, for \textit{M}-cells by equation \eqref{eq:migrationFun-M}. Parameters are the migration rate constants $c_A$ and $c_M$, which are chosen $c_A=1$ and $c_M=0.5$ in the plot. }
\label{fig:rates}
\end{figure}

\begin{figure}[ht]
\centering
\includegraphics[width=\linewidth]{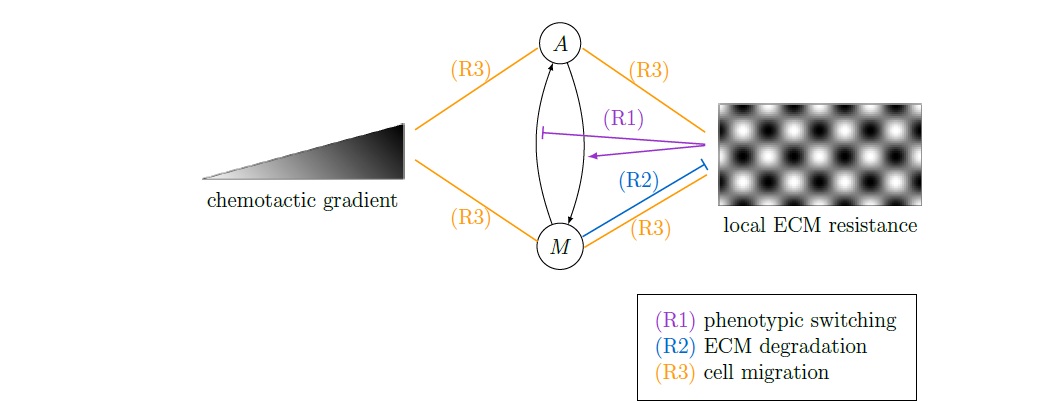}
\caption{Schematic illustration of model rules. The time evolution of our CA model arises from the repeated application of three rules: (R1) phenotypic switching between \textit{A}- and \textit{M}-cells depending on ECM resistance, (R2) ECM degradation by \textit{M}-cells and (R3) cell migration depending on the ECM resistance, preferentially in direction of the chemotactic gradient. The black arrows indicate the switch ability between \textit{A}- and \textit{M}-cells. Relations between chemotactic gradient, \textit{A}/\textit{M}-cells and local ECM resistance are indicated by arrows (positive influence), stopped lines (negative influence) and straight lines (positive or negative influence).  }
\label{fig:modelScheme}
\end{figure}

The time evolution of our model is defined by the following rules:

\hspace*{0.1em}\parbox{0.96\textwidth}{\renewcommand{\labelenumi}{\textbf{(A\theenumi)}}\begin{enumerate}\setlength{\itemsep}{0.5em}
	\item[(R1)] \textit{Phenotypic switch}. A cell changes its phenotype depending on the local ECM resistance $\mu$. In particular, an \textit{A}-cell changes its migratory phenotype with rate $\alpha\mu$ and an \textit{M}-cell with rate $\beta(1-\mu)$, respectively, see Figure \ref{fig:rates}(a).
	\item[(R2)] \textit{ECM degradation}. \textit{M}-cells locally degrade the ECM with constant rate $\delta>0$.	
	\item[(R3)] \textit{Cell migration}. \textit{A}-cells migrate with rate $\lambda_A(\mu)$, \textit{M}-cells with rate $\lambda_M(\mu)$, where both $\lambda_A(\mu)$ and $\lambda_M(\mu)$ depend on the local ECM resistance $\mu$, see Figure \ref{fig:rates}(b). The direction of movement is determined by a chemotactic gradient. Cells respond to this gradient with an intensity $\kappa>0$. If $\kappa=0$, cells perform independent random walks, which is equivalent to the case when no gradient field is present. 	
\end{enumerate}	} \\[0.25em]	
At each discrete model time step, one cell is selected at random. The state of the selected cell is updated according to the rules (R1)-(R3), where cell migration is implemented as an exclusion process \cite{Simpson2009}. The latter means that cells move to neighboring nodes if the target node is empty, otherwise the movement is aborted. 
Thus, the model accounts for steric interactions between cells due to volume exclusion.
A detailed description of the model update rules can be found in Supplementary Section 1. A schematic illustration of the update rules is depicted in Figure \ref{fig:modelScheme}.

In each simulation, we measure the migration distance from the initial position for each individual cell ($d$). At the population level, the model observable characterizing invasion dynamics is the migration distance from the initial position averaged over the entire cell population ($d_p$). In addition, we consider different simulations to compare $d_p$-values between populations of cells where each cell is allowed to switch between ameoboid and mesenchymal migration modes and those of non-switching cell populations ($\Delta d_{max}$). 
Calculation details are given in Supplementary Section 2.

\subsection*{Simulation study}

\begin{figure}[ht]
\centering
\includegraphics[width=\linewidth]{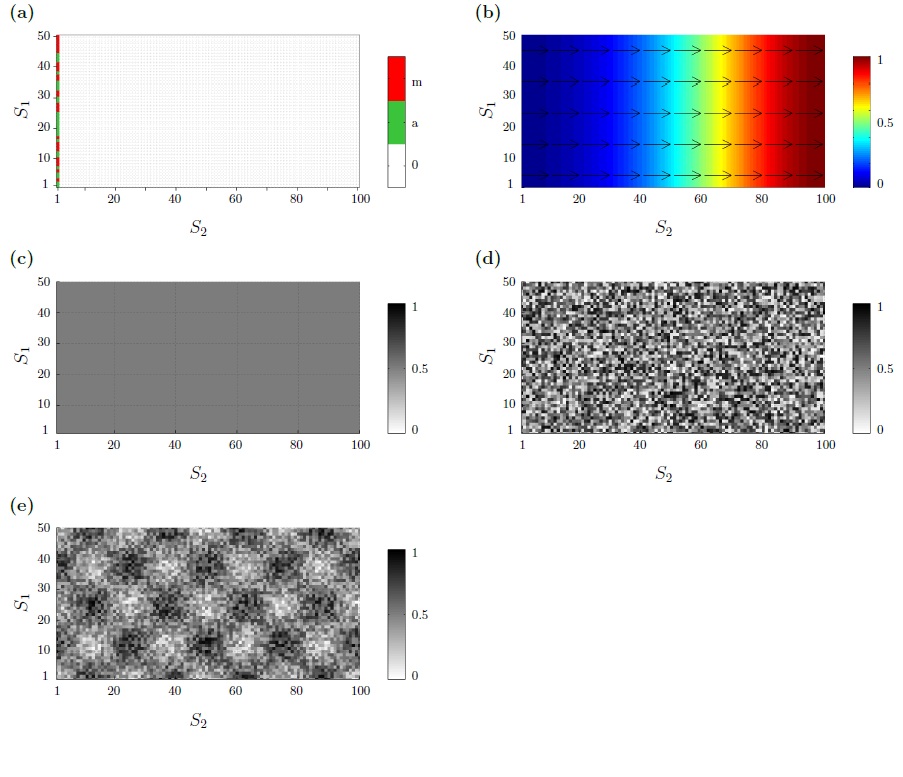}
\caption{Initial model conditions. 
	(a) Initial cell distribution. \textit{A}- and \textit{M}-cells are placed at random along the left border of a rectangular lattice. Periodic boundary conditions are applied along the $S_1$ axis. At $S_2=1$ reflecting boundary conditions are imposed. 
	(b) Chemotactic gradient field defined by equation \eqref{eq:gradientFun}. The color code indicates the chemotactic gradient concentration function, where blue corresponds to low, red is related to high concentrations. The arrows illustrate the local gradient direction.
	(c)-(e) Initial ECM resistance distribution of differently structured ECM.
	(c) homogeneous, moderate ECM resistance distribution, modeled by a constant value $\mu(r_1,r_2)=0.5$. 
	(d) heterogeneous, weakly structured ECM resistance distribution, defined by equation \eqref{eq:ECMFun} with heterogeneity parameter $\theta=0.1$; 
	(e) heterogeneous, highly structured ECM resistance distribution, defined by equation \eqref{eq:ECMFun} with heterogeneity parameter $\theta=0.5$;
	The gray scale in (c)-(e) indicates the level of matrix resistance on cell migration: low (white) to high (black).		 }
\label{fig:IC}
\end{figure}

The CA model is simulated on a rectangular lattice with dimensions $S_1\times S_2$. The vertical length of the lattice is given by the $r_1$ coordinate, $1\leq r_1\leq S_1$ and the horizontal length of the lattice is given by the $r_2$ coordinate, $1\leq r_2\leq S_2$. For the simulations we chose initial and boundary conditions such that we can analyze cell migration using one-dimensional, vertically averaged, cell density profiles. Periodic boundary conditions are applied along the horizontal boundaries ($r_1=1$ and $r_1=S_1$), and reflecting boundary conditions on the vertical boundaries ($r_2=1$ and $r_2=S_2$). 
The initial conditions of the model are illustrated in Figure \ref{fig:IC}. At the initial time, 50\% \textit{M}-cells and 50\% \textit{A}-cells are placed at random along the left border at $r_2=1$, as depicted in Figure \ref{fig:IC}(a). Notice that the results do not depend on the initial fractions of \textit{M}- and \textit{A}-cells but on the phenotypic switch ratio $\alpha/\beta$ (simulations not shown).
Figure \ref{fig:IC}(b) shows the chemotactic gradient field, which controls the direction of cell movement. The chemotactic gradient is chosen such that the one-dimensional cell density profile advances along the $S_2$-axis. In particular, the gradient is specified by a linear function
\begin{equation}\label{eq:gradientFun}
	G(r_1,r_2)=\frac{1}{S_2}r_2,
\end{equation}
which describes an attractive signal from the target location $r_2=S_2$, see Figure \ref{fig:IC}(b). We consider different chemotactic responsiveness parameters $\kappa\geq 0$, reflecting a range from undirected cell movement ($\kappa=0$) to a high response to the presence of a chemotactic gradient ($\kappa \gg 0$).

For the ECM resistance, different initial scenarios are implemented. Homogeneous ECM is modeled by a constant ECM resistance, as illustrated in Figure \ref{fig:IC}(c). 
Heterogeneous ECM is modeled by a sinusoidal function
\begin{equation}\label{eq:ECMFun}
	\mu(r_1,r_2)= 1-\frac{\theta}{2} + \frac{\theta}{2}\sin\left(\frac{4\pi}{S_1}r_1+\frac{3\pi}{2}\right)\sin\left(\frac{8\pi}{S_2}r_2 + \frac{\pi}{2}\right)  - (1-\theta)\xi,
\end{equation}
where $\xi\sim\mathcal{U}(0,1)$ is a fixed uniformly distributed random variable and $\theta\in[0,1]$ is a heterogeneity parameter. 
On average, the ECM resistance is $\mu(r_1,r_2)\approx 0.5$.
The heterogeneity parameter controls the level of ECM structure, ranging from completely random ($\theta=0$) to completely coherent structures ($\theta=1$). An example of a less structured ECM is shown in Figure \ref{fig:IC}(d), a highly structured ECM is illustrated in Figure \ref{fig:IC}(e). 
We have also tested other functional representations of the ECM resistance distribution. The results we describe below remain valid for other functional forms of $\mu$, as long as $\mu$ is not a constant but modeled by a spatial function $\mu:S_1\times S_2\rightarrow [0,1]$ that depends on the spatial position $(r_1,r_2)$ and on a random component, where the latter adds local irregularities to an otherwise regular spatial structure. See Supplementary Fig. S1  for further examples.

The \textit{A}- and \textit{M}-cell migration rates are modeled by sigmoidal functions
\begin{subequations}
\begin{align}
	\lambda_A(\mu) &=c_A/(1+\exp(15(\mu-0.5))) 	\label{eq:migrationFun-A}\\
	\lambda_M(\mu) &= c_M/(1+\exp(15(\mu-0.5))) \label{eq:migrationFun-M},
\end{align}
\end{subequations}
where $c_A$ and $c_M$ are constants, chosen such that the migration speed of \textit{A}-cells is higher than that of \textit{M}-cells for low to moderate ECM resistance, see Figure \ref{fig:rates}(b). For $\mu=0.5$, the slopes of the migration rate functions are steepest. 
The sigmoidal shape of the migration rate functions accounts for three levels of matrix resistance on cell migration: low, middle and high. Cell migration is unhindered for low ECM resistance, linearly decreases for medium ECM resistance and is impeded for high ECM resistance.

The focus of our study is on the impact of the switch parameters $\alpha$ and $\beta$ on the migration behavior of the cell population. We refer to a \textit{switching cell population} if cells are allowed to change their phenotype ($\alpha,\beta>0$), and to a \textit{non-switching cell population} if cells do not adapt their phenotype ($\alpha=\beta=0$). 
For switching populations, we change the phenotypic switch parameters $\alpha=1/10,\dots,1$, $\beta=1$ and $\alpha=1$, $\beta=1/10,\dots,1$, such that the phenotypic switch ratio is $\alpha/\beta\in\{1/10,2/10,\dots,1,2,\dots,10\}$. 
We measure how far the switching population migrates from the start position. Notice that this average distance does not depend on the particular choice of $\alpha$ and $\beta$ as long as the switch ratio $\alpha/\beta$ remains fixed (simulations not shown). 
In order to identify novel behaviors under the influence of the phenotypic switching, we compare our results with the non-switching situation $\alpha=\beta=0$, that is when \textit{A}- and \textit{M}-cells do not change their phenotype. For the non-switching population, we simulate the same number of cells as in the switching scenario, with changing ratio of \textit{A}- and \textit{M}-cells.  The fraction of \textit{M}-cells in a non-switching population is denoted by $\gamma$.
Notice that the non-switching situation reflects a modeling scenario similar to the one studied by Hecht et al.~\cite{Hecht2015}. 

In our simulations, we compare the migration behavior of switching and non-switching cell populations under different environmental conditions: homogeneous or heterogeneous ECM structure in combination with different chemotactic responsiveness. A summary of all possible simulation scenarios is given in Table \ref{tab:simScenarios}. In each scenario, we compare the migration distance $d_p$ of switching and non-switching populations of size $n=50$. 
For the switching population, we change the phenotypic switch ratio $\alpha/\beta$ as described above. For the non-switching population, we vary the fraction of \textit{M}-cells:  $\gamma=1$ (pure \textit{M}-cell population), $\gamma=0.7$ (70\% \textit{M}-cells, 30\% \textit{A}-cells), $\gamma=0.3$ (30\% \textit{M}-cells, 70\% \textit{A}-cells) and $\gamma=0$ (pure \textit{A}-cell population). 
Each simulation is performed for 200 Monte Carlo steps\footnote{A Monte Carlo step is the standard time unit in our system. In particular, let $n$ be the total number of cells in our model. At each discrete time step, one cell is selected at random and updated according to our model rules. On average, each cell is updated once per $n$ consecutive simulation time steps. The sequence of $n$ time steps corresponds to one Monte Carlo step.}.
Throughout the study, we change the migration rate ratio $c_M/c_A$ by varying the \textit{M}-cell migration rate ($0\leq c_M\leq 1$), while keeping the \textit{A}-cell migration rate fixed ($c_A=1$). A small ratio $c_M/c_A$ indicates a large difference between the two migration rates, if the ratio $c_M/c_A$ is close to 1, \textit{M}-cells migrate nearly as fast as \textit{A}-cells. The ECM degradation rate is set to $\delta=0.1$ ensuring partial but not complete degradation. An overview of the model parameters is given in Table \ref{tab:modelParameters}. 

As a final step, we study if cooperative effects in a switching population contribute to an efficient migration of the entire cell population.
To this end, we compare the migration behavior of a single cell moving alone with that of an individual cell moving in a cell population of size $n=500$. In particular, we first simulate $n$ repetitions of the single cell scenario, each time measuring the migration distance $d$ of the single cell, and at the end determine the maximum migration distance $d$. Secondly, we simulate a cell population of size $n$, using the same model parameters as in the single-cell scenario. We measure the migration distance $d$ of each individual cell of the population and determine the maximum migration distance $d$. Both simulations are repeated 100 times to account for stochastic fluctuations.

\section*{Results}

\subsection*{Advantage of non-switching behavior under homogeneous ECM conditions}
 
\begin{figure}[ht]
\centering
\includegraphics[width=\linewidth]{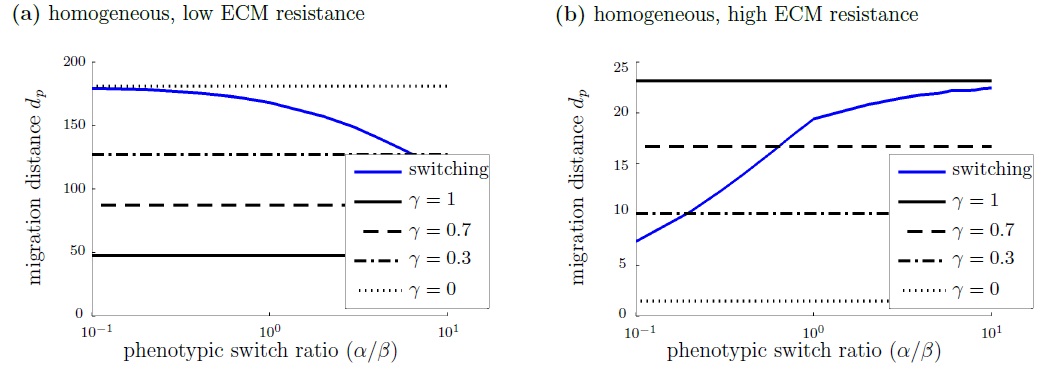}
\caption{Advantage of non-switching behavior under homogeneous ECM conditions. 
	The figure shows the migration distance $d_p$ of switching and non-switching populations depending on the switch ratio $\alpha/\beta$ for different homogeneous ECM conditions: (a) homogeneous, low ECM resistance modeled by $\mu(r_1,r_2)=0.1$ and (b) homogeneous, high ECM resistance with $\mu(r_1,r_2)=0.9$. The blue line represents the switching populations, the black lines display non-switching populations with different \textit{M}-cell fraction $\gamma\in\{0,0.3,0.7,1\}$. Each simulation is run with $50$ cells. Simulations are evaluated after 200 Monte Carlo steps, averaged over 50 independent simulations. The standard deviation is not shown as it is negligible. Simulation parameters are $c_M/c_A=0.25$, $\kappa=1$, $\delta=0.1$. }
\label{fig:influenceHOM}
\end{figure}

We first study if phenotypic switching provides an advantage in terms of cell population migration distance under homogeneous ECM conditions. For this, we compare the migration distance $d_p$ of switching and non-switching populations under homogeneous ECM conditions, varying the level of $\mu(r_1,r_2)=constant$ initially, in combination with a high chemotactic responsiveness.
Figure \ref{fig:influenceHOM} illustrates simulation results under different homogeneous ECM resistance conditions: homogeneous, low ECM resistance (Figure \ref{fig:influenceHOM}(a)) and homogeneous, high ECM resistance (Figure \ref{fig:influenceHOM}(b)). The figure shows the migration distance $d_p$ of the switching cell population as a function of the phenotypic switch ratio $\alpha/\beta$. Also shown is the migration distance $d_p$ of non-switching populations with different \textit{M}-cell fraction $\gamma\in\{0,0.3,0.7,1\}$, which remains constant as a function of $\alpha/\beta$.
In Figure \ref{fig:influenceHOM}(a), for homogeneous, low ECM resistance, we observe that the higher the \textit{M}-cell fraction $\gamma$ of the non-switching population, the smaller the migration distance $d_p$. 
Similarly, the migration distance $d_p$ of the switching population is highest for low switch ratio $\alpha/\beta$, which means that cells predominantly switch to \textit{A}-type. As the phenotypic switch ratio $\alpha/\beta$ increases, that is fewer \textit{A}-cells and more \textit{M}-cells are present, the migration distance $d_p$ of the switching population decreases as the phenotypic switch ratio $\alpha/\beta$ increases.
In contrast, under homogeneous, high ECM conditions, a non-switching population with
high \textit{M}-cell fraction $\gamma$ shows a higher migration distance $d_p$ than a non-switching population with small \textit{M}-cell fraction, see Figure \ref{fig:influenceHOM}(b). Likewise, for the switching population we observe that increasing the switch ratio $\alpha/\beta$, which leads to a preferred switch to \textit{M}-type cells, increases the migration distance $d_p$. 
Under both homogeneous ECM conditions, low and high, we find that the possibility to switch the phenotype does not constitute a benefit in terms of migration distance of the entire cell population. 
The critical ECM resistance value above which the greatest migration distance is no longer observed for the pure \textit{A}-cell population but for the pure \textit{M}-cell population, is approximately $\mu(r_1,r_2)=0.5$. 

The observed behavior can be understood by considering the cell migration characteristics. On the one hand, for low to moderate, homogeneous ECM resistance, cell migration is not hindered and \textit{A}-cells migrate with higher rate than \textit{M}-cells. Thus, the higher the fraction of \textit{A}-cells is, the higher the migration distance $d_p$ of the entire cell population. On the other hand, sufficiently high homogeneous ECM resistance presents a barrier for \textit{A}-cells and only \textit{M}-cells are able to create space to migrate through the ECM. Therefore, a high \textit{M}-cells fraction increases the migration distance $d_p$.

\subsection*{Migration plasticity can be advantageous under heterogeneous ECM conditions}

\begin{figure}[ht]
\centering
\includegraphics[width=\linewidth]{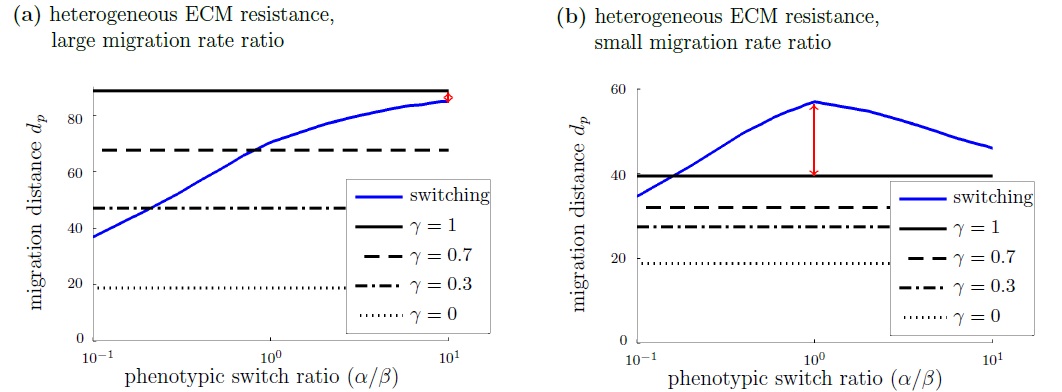}
\caption{Migration plasticity can be advantageous under heterogeneous, highly structured ECM conditions.
	The figure shows the migration distance $d_p$ of switching and non-switching populations depending on the switch ratio $\alpha/\beta$ under heterogeneous ECM condition (modeled by equation \eqref{eq:ECMFun} with $\theta=0.5$, see Figure \ref{fig:IC}(e)) and with different migration rate ratios: (a) $c_M/c_A=0.75$ and (b) $c_M/c_A=0.25$. The blue line represents the switching populations, the black lines display non-switching populations with different \textit{M}-cell fraction $\gamma\in\{0,0.3,0.7,1\}$. Each simulation is run with $50$ cells. Simulations are evaluated after 200 Monte Carlo steps, averaged over 50 independent simulations. The standard deviation is not shown as it is negligible. Simulation parameters are $\kappa=1$, $\delta=0.1$. 
	The red arrow indicates the difference  $\Delta d_{max}$ between the switching population with maximum migration distance $d_p$ with respect to varied $\alpha/\beta$ ratio and the non-switching population with maximum $d_p$ with respect to varied $\gamma$ fraction, which is the observable analyzed in Figure \ref{fig:phaseDiagram}. }
\label{fig:influenceHET}
\end{figure}

In a next step, we determine if a heterogeneous ECM structure provides an environmental condition under which the switching population of cells migrates the greatest distance from the initial position. To this end, we compare the migration distance $d_p$ of switching and non-switching populations under heterogeneous ECM conditions ($\mu$ as defined in equation \eqref{eq:ECMFun}), varying the levels of ECM heterogeneity $\theta$, in combination with a high chemotactic responsiveness.
Under heterogeneous, weakly structured ECM conditions, we observe that the maximum migration distance $d_p$ is obtained by the non-switching population.
In contrast, under heterogeneous, highly structured ECM conditions, we find that the migration distance $d_p$ of the switching population may be lower or higher than of the non-switching population, depending on the migration rate ratio $c_M/c_A$. Simulation results are illustrated in Figure \ref{fig:influenceHET}. 
Figure \ref{fig:influenceHET}(a) shows the migration distance $d_p$ of switching and non-switching populations as a function of the switch ratio $\alpha/\beta$ under heterogeneous, highly structured ECM conditions and a large migration rate ratio ($c_M/c_A\approx 1$). We observe that the higher the portion of \textit{M}-cells, the higher the migration distance $d_p$ of the non-switching population. Similarly, for the switching population, the higher the switch ratio $\alpha/\beta$, that is cells are predominantly \textit{M}-type, the higher the migration distance $d_p$. The non-switching cell population with $\gamma=1$ (pure \textit{M}-cell population) migrates the greatest distance from the initial position.
The situation is different if the migration rate ratio is small ($c_M/c_A\ll 1$): while the migration distance $d_p$ of the non-switching population is higher, the more \textit{M}-cells are present, the maximum migration distance $d_p$ of the switching population is obtained for $\alpha/\beta=1$, that is when both cell types are equally present, see Figure \ref{fig:influenceHET}(b). 
Under the latter conditions, that is the ECM is heterogeneous, highly structured and the difference between \textit{A}- and \textit{M}-cell migration speeds is large, we observe that the switching behavior provides an advantage in terms of cell population migration distance.

\begin{figure}[ht]
\centering
\includegraphics[width=\linewidth]{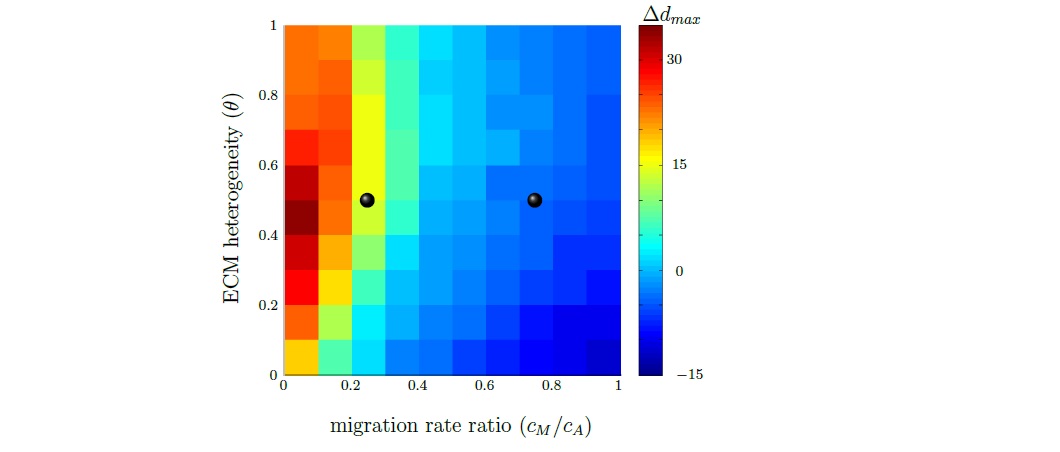}
\caption{The difference $\Delta d_{max}$ between the maximum migration distance $d_p$ of switching and non-switching populations depends on the ECM heterogeneity ($\theta$) and the migration rate ratio ($c_M/c_A$). The phase diagram shows the difference $\Delta d_{max}$ between the switching population with maximum migration distance $d_p$ with respect to varied switch ratio $\alpha/\beta$ and the non-switching population with maximum migration distance $d_p$. Red to turquois areas indicate parameter combinations $(\theta,c_M/c_A)$ for which the maximum migration distance $d_p$ of the switching population is highest, whereas blue refers to an advantage of the non-switching behavior. The black dots illustrate points in the parameter space, which corresponds to the parameter situation of Figures \ref{fig:influenceHET}(a) and (b). Simulation parameters are $\kappa=1$, $\delta=0.1$. }
\label{fig:phaseDiagram}
\end{figure}

Figure \ref{fig:phaseDiagram} shows the observed dependency of the  migration distance $d_p$ on the level of ECM heterogeneity and the cell migration rate ratio. In the phase diagram, the difference between the switching population with maximum migration distance $d_p$ with respect to varied switch ratio $\alpha/\beta$ and the non-switching population with maximum $d_p$ is illustrated. 
We observe that the switching behavior is favorable if the ECM is highly structured and the migration rate ratio is small. Otherwise the maximum migration distance $d_p$ is reached by the non-switching population. 
We conclude that, provided the difference between \textit{A}- and \textit{M}-cells is sufficiently large enough, a high chemotactic responsiveness together with a pronounced physical resistance imposed by the ECM, represents an environmental situation under which phenotypic adaptation of cell migration modes is beneficial. We hypothesize that a chemotactic gradient causes directional movement of cells but the locally favorable direction may not be the overall best choice. While rather fast \textit{A}-cell movement is impeded by high ECM resistance, slower moving \textit{M}-cells create their own short paths. Thus, if cells in the population are allowed to adapt their phenotype to the local environmental conditions, they are able to follow the shortest path towards the target location and to overcome local ECM constraints.

\subsection*{Chemotactic responsiveness influences the efficiency of cell movement}

We now compare the migration behavior of switching and non-switching populations under different chemotactic responsiveness. To determine if the response towards the chemotactic gradient alters the simulation results described in the previous sections, we repeat the simulation study where we explored homogeneous and heterogeneous ECM conditions, this time however changing the chemotactic responsiveness.
The result of this sensitivity analysis is that under homogeneous ECM conditions, as well as under heterogeneous, weakly structured ECM conditions, the migration distance $d_p$ of the non-switching population is higher than of the switching population, independent of the chemotactic responsiveness, as shown in Supplementary Fig. S2 and Supplementary Fig. S3. 
Under heterogeneous, highly structured ECM conditions, we find that the advantage of phenotypic switching in terms of cell population migration distance depends on the chemotactic responsiveness, such that the advantage of the switching behavior becomes more pronounced the higher the chemotactic responsiveness. In particular, if the ECM is highly structured and the migration rate ratio $c_M/c_A$ is small, the difference between the migration distance $d_p$ of switching and non-switching populations increases with increasing chemotactic responsiveness, see Supplementary Fig. S4.

\subsection*{Cooperative effects increase the migration distance}

\begin{figure}[ht]
\centering
\includegraphics[width=\linewidth]{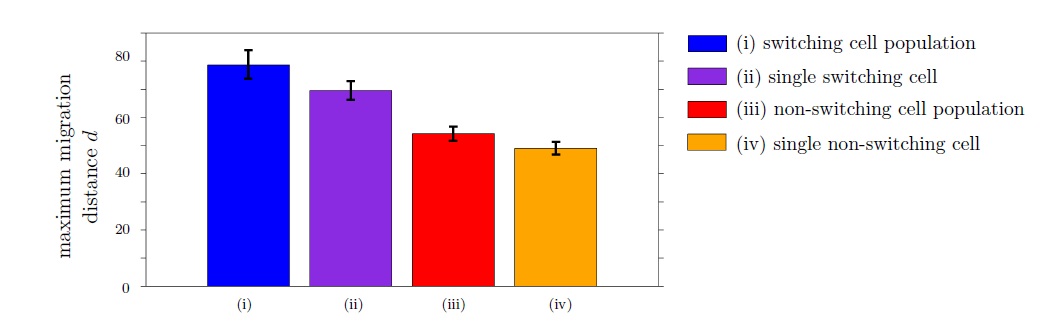}
\caption{An individual cell within a switching population can migrate further compared to a situation where only a single switching cell is moving.
CA simulations of (i) a switching population of 500 cells, (ii) a scenario with only a single switching cell repeated 500 times, (iii) a non-switching population of 500 cells and (iv) a scenario with only a single non-switching cell repeated 500 times. 
The figure shows the maximum migration distance $d$, which refers to the maximum distance the individual cells within a population migrate from the initial position (i, iii), or to the maximum distance the single cell migrates among the 500 repetitions (ii, iv), respectively.
Each simulation, (i)-(iv), is performed for 200 Monte Carlo steps and repeated 100 times to account for stochastic fluctuations. The ECM condition and model parameters are chosen as in Figure \ref{fig:influenceHET}, with phenotypic switch ratio $\alpha/\beta=1$.}
\label{fig:singleVsMulti}
\end{figure}

To study if cooperative effects in a switching population influence the efficiency of cell movement, we measure the maximum migration distance $d$ of the switching population for different population sizes. In particular, the maximum distance $d$ a single switching cell migrates among 500 simulation repetitions is compared to the migration distance $d$ of the farthest individual cell in a switching cell population of size $n=500$, as described in the Methods section. 
Figure \ref{fig:singleVsMulti} illustrates simulation results under heterogeneous, highly structured ECM conditions. We observe that the maximum migration distance $d$ is higher for a switching cell population than for a single cell which adapts its migratory phenotype. For comparison, we repeat the simulation procedure for the non-switching situation, with results also shown in Figure \ref{fig:singleVsMulti}. We find that the maximum migration distance $d$ of the non-switching population is lower than the switching situation, and lowest in the case of a single non-switching cell. 
We conclude that cooperative effects are present in a switching population. We hypothesize that \textit{M}-cells benefit \textit{A}-cells by creating space in an otherwise impenetrable ECM. Our argument is in agreement with the findings of Hecht et al. \cite{Hecht2015}, who showed that in a non-switching-population, paths created by mesenchymal cells are used by amoeboid cells. In fact, our results show that cooperative effects are present in a switching population and allow the cells to migrate longer distances compared to the non-switching situation.

\section*{Discussion}

We developed a cellular automaton model to study the impact of amoeboid-mesenchymal migration plasticity on tumor invasion. The switch between the two migration modes was assumed to depend on the local microenvironment through the biomechanical resistance imposed by the ECM. Cells in the model are able to switch between a slow mesenchymal migration mode with ECM degradation, and a fast amoeboid migration mode without ECM degradation, depending on ECM resistance. 
With computer simulations of the mathematical model we analyzed invasion dynamics, characterized by the migration distance of the cell population, under different environmental conditions. In particular, we distinguished spatially homogeneous and heterogeneous ECM resistance conditions, ranging from weakly structured to highly structured, in combination with different responses towards a chemotactic gradient which directs cell migration. Provided that the difference between amoeboid and mesenchymal migration speeds is sufficiently large, as observed in experiments \cite{Liu2015}, we found that an amoeboid-mesenchymal migration plasticity provides an advantage in terms of migration distance if the spatial structure of ECM resistance is strongly heterogeneous and if migration is directed.

In our model, we only distinguish two migration modes, which we term amoeboid and mesenchymal. The mesenchymal mode is characterized by slow migration speed and ECM degradation, while cells in amoeboid mode move fast and do not degrade the ECM. It is clear that this is a strong simplification of biological reality. Cells may display features of both migration modes, in particular concerning parameters that are not explicitly incorporated in our model, such as Rho/ROCK signaling and cell shape changes. However, our coarse-grained model captures the extreme cases of plasticity behavior. Therefore, we expect that our predictions remain valid also for intermediate situations.

In this paper, we assumed that physical and biomechanical properties of the ECM, such as density and stiffness, can be combined in a single parameter called ECM resistance. We are aware that this simplification does not account for the structural and molecular complexity of the ECM. However, we coarse-grain the extra- and intracellular details into a simplified model to provide a basic understanding of emerging plasticity effects. Our model is a first step to analyze consequences of migration plasticity on tumor invasion.

The chemical gradient which directs cell migration in the model is assumed to be independent of the ECM resistance. It is known that chemotaxis is important for tumor invasion \cite{Roussos2011}. Other types of directed cell migration have been observed in tumor cells, such as haptotaxis (movement along gradients of ECM ligands) or durotaxis (gradient of ECM stiffness). Our choice to consider ECM-independent chemotaxis serves as a first attempt to incorporate the effects of directed migration into an amoeboid-mesenchymal plasticity model for tumor invasion. We expect that our results hold for other than chemotactic gradients.

In the model, we assumed that amoeboid and mesenchymal cell migration speeds decrease monotonically with increasing ECM resistance. We based our assumption on findings of Wolf et al. \cite{Wolf2013}, where it was shown, by using breast cancer and fibrosarcoma cell lines, that cell migration speed does decrease monotonically with decreasing ECM pore size but does not depend on ECM stiffness. Both pore size and ECM stiffness contribute to the biomechanical ECM resistance against cell migration. However, experimental findings on the relation between cell migration speed and the ECM environment are controversial. For example, experiments with brain tumor cell lines have shown that an increase of  ECM stiffness induces an increased cell migration speed  \cite{Ulrich2009}. A possible explanation for different experimental findings might be the type of tumor studied. Furthermore, it is well-known that while ECM porosity and stiffness can be independently controlled under \textit{in vitro} conditions, \textit{in vivo}, stiffening of the ECM is highly correlated with alterations of the ECM fiber structure \cite{Das2013}. Clarifying the particular contributions of different ECM properties to cell migration is primarily an experimental task.

In our model, we indirectly account for energetic costs of migration plasticity. Different cellular tasks as migration and matrix degradation consume energy and we assume a trade-off between the energies invested in these tasks. Consequently, cells in mesenchymal migration mode which degrade the ECM move slower than amoeboid cells which do not spend energy for proteolysis. This assumed difference in migration speeds is supported by several biological findings \cite{Liu2015,Wolf2003}. Our results on the existence of cooperative effects in switching cell populations support previous findings of Hecht et al. \cite{Hecht2015}. They showed that an amoeboid cell population can benefit from the presence of a small number of path creating mesenchymal cells in a model where metabolic cost for ECM degradation is modeled directly as trade-off with migration speed.

Previous theoretical works have studied amoeboid and mesenchymal cell migration under different environmental conditions \cite{Tozluoglu2015,Hecht2015}. Here, for the first time, the impact of plasticity of amoeboid and mesenchymal migration modes on tumor invasion is investigated. Contrary to our initial expectation, amoeboid-mesenchymal migration plasticity is not advantageous per se. Interestingly, it is only with the combined influence of ECM heterogeneity and chemotactic responsiveness that the switching behavior provides an advantage in terms of migration distance. So far, in experimental studies the focus was either on heterogeneous ECM structure or on varying chemotactic responsiveness. However, it is well-known that in real tumors the ECM is characterized by structural heterogeneity \cite{Erler2009,Polacheck2013} and the presence of chemotactic gradients at the same time.  \cite{Roussos2011,MuinonenMartin2014}. 
Our study suggests that in the case of complex heterogeneous environments accompanying tumor invasion, the ability of tumor cells to switch between amoeboid and mesenchymal modes is a trigger for rapid cell population invasion.

\bibliography{migrationStrategy}

\section*{Acknowledgements}

The authors would like to thank Katarina Wolf (Nijmegen, Netherlands) and Miguel A. Herrero (Madrid, Spain) for helpful discussions.
AD acknowledges support by Deutsche Krebshilfe.
AD and EACA were financially supported by SFB TRR79 projects M8 an B5.
AVB acknowledges support by S{\"a}chsisches Staatsministerium f{\"u}r Wissenschaft und Kultur (SMWK) in the framework of INTERDIS-2.
The authors thank the Center for Information Services and High Performance Computing at TU Dresden for providing an excellent infrastructure.

\section*{Author contributions statement}

Wrote the paper and discussed results: KB EACA AD AVB. Supervised the study and gave substantial input to the manuscript: AD AVB. Conceived and designed the model and performed simulations: KB. Analyzed the model: KB AVB. All authors reviewed the manuscript. 

\section*{Additional information}

\textbf{Competing financial interests:} The authors declare no competing financial interests.

The corresponding author is responsible for submitting a \href{http://www.nature.com/srep/policies/index.html#competing}{competing financial interests statement} on behalf of all authors of the paper. This statement must be included in the submitted article file.

\begin{table}[ht]
\centering
\renewcommand{\arraystretch}{1.5}
	\begin{tabular}{|cL{3cm}C{2.8cm}|C{1.75cm}|C{1.76cm}|C{1.75cm}|}\cline{4-6}
	\multicolumn{3}{c|}{\multirow{2}{*}{}}&\multicolumn{3}{c|}{chemotactic responsiveness ($\kappa$)}\\
	\multicolumn{3}{c|}{}& none & low & high \\[-0.8em]
	\multicolumn{3}{c|}{}& ($\kappa=0$) & ($\kappa=0.01$) & ($\kappa=1$) \\\hline
	\multirow{8}{*}{\rotatebox{90}{ECM ($\mu$)}}& 
		\multirow{4}{3cm}{homogeneous \newline ($\mu=constant$)} &low resistance&
	\multicolumn{3}{c|}{\multirow{9}{6cm}{\includegraphics[width=0.35\textwidth]{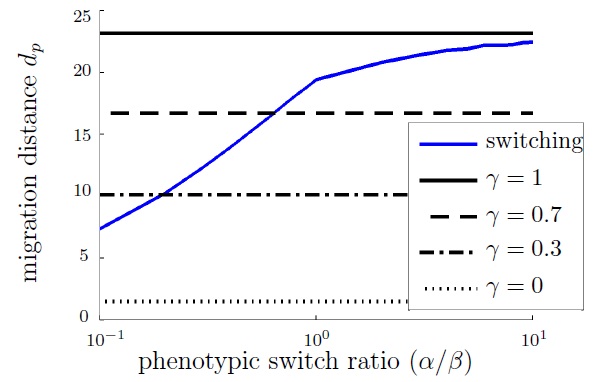} }}\\[-0.6em]
	&&($\mu=0.1$)		&\multicolumn{3}{c|}{}\\\cline{3-3}
	& & high resistance	&\multicolumn{3}{c|}{}\\[-0.6em]
	&& ($\mu=0.9$)		&\multicolumn{3}{c|}{}\\
	\cline{2-3}

	&\multirow{4}{3cm}{heterogeneous \newline($\mu$ see equation (3) main text, with heterogeneity parameter $\theta$)} &weakly structured &\multicolumn{3}{c|}{}\\[-0.6em]
	& &($\theta=0.1$)&\multicolumn{3}{c|}{}\\
	\cline{3-3}
	& &highly structured&\multicolumn{3}{c|}{}\\[-0.6em]
	& &($\theta=0.5$)&\multicolumn{3}{c|}{}\\[0.5em]
	\hline
	\end{tabular}
\renewcommand{\arraystretch}{1}
\caption{Overview of simulation scenarios. We study different environmental conditions: homogeneous or heterogeneous ECM structure (left column) in combination with different chemotactic responsiveness (top row). In each scenario, we measure the migration distance $d_p$ of switching and non-switching populations depending on the switch ratio $\alpha/\beta$. As illustrated in the figure, the switching population results (blue line) are compared to those of non-switching populations with different \textit{M}-cell fractions: $\gamma=1$ (pure \textit{M}-cell population; solid line), $\gamma=0.7$ (70\% \textit{M}-cells, 30\% \textit{A}-cells; dashed line), $\gamma=0.3$ (30\% \textit{M}-cells, 70\% \textit{A}-cells; dash-dot line) and $\gamma=0$ (pure \textit{A}-cell population; dotted line).}
\label{tab:simScenarios}
\end{table}

\begin{table}[ht]
\centering
	\begin{tabular}{|c|p{0.55\textwidth}|c|}
		\hline
		\textbf{Parameter} & \textbf{Description} & \textbf{Value}  \\
		\hline\hline
		$\alpha $ 	& phenotypic switch rate of an \textit{A}-cell to become an \textit{M}-cell per unit time & 0.1 to 1\\
		$\beta $ 	& phenotypic switch rate of an \textit{M}-cell to become an \textit{A}-cell per unit time& 0.1 to 1\\
		$c_A$		& \textit{A}-cell migration rate constant & 1\\
		$c_M$		& \textit{M}-cell migration rate constant & 0 to 1\\		
		$\delta $ 	& ECM degradation rate per unit time& 0.1 \\
		$\gamma$	& fraction of \textit{M}-cells in non-switching population & 0 to 1\\
		$\kappa $ 	& chemotactic responsiveness & 0 to 1 \\
		$\theta$	& ECM heterogeneity parameter	 & 0 to 1\\
		\hline
	\end{tabular}
\caption{\label{tab:modelParameters} Overview of all model parameters.}
\end{table}

\end{document}